\documentclass[aps,floats,showpacs,epsf,superscriptaddress,twocolumn]{revtex4}
\DeclareMathAlphabet{\mathpzc}{OT1}{pzc}{m}{it} 
\usepackage{amsfonts, color}
\usepackage{graphics}
\usepackage{graphicx}

\begin{document}

\title{Theory of integer quantum Hall effect in insulating bilayer graphene}
\author{Bitan Roy}

\affiliation{National High Magnetic Field Laboratory, Florida State University, Florida 32306, USA}
\affiliation{Condensed Matter Theory Center, Department of Physics, University of Maryland, College Park, MD 20742, USA}

\date{\today}

\begin{abstract}
A variational ground state for insulating bilayer graphene (BLG), subject to quantizing magnetic fields, is proposed. Due to the Zeeman coupling, the layer anti-ferromagnet (LAF) order parameter in fully gapped BLG gets projected onto the spin easy plane, and simultaneously a ferromagnet order, which can further be enhanced by exchange interaction, develops in the direction of the magnetic field. The activation gap for the $\nu=0$ Hall state then displays a crossover from quadratic to linear scaling with the magnetic field, as it gets stronger, and I obtain excellent agreement with a number of recent experiments with realistic strengths for the ferromagnetic interaction. A component of the LAF order, parallel to the external magnetic field, gives birth to additional incompressible Hall states at filling $\nu=\pm 2$, whereas the remote hopping in BLG yields $\nu=\pm 1$ Hall states. Evolution of the LAF order in tilted magnetic fields, scaling of the gap at $\nu=2$, the effect of external electric fields on various Hall plateaus, and different possible hierarchies of fractional quantum Hall states are highlighted.        
\end{abstract}


\maketitle

\vspace{10pt}

 Two dimensional chiral electron gas in single and bilayer graphene respectively discerns anomalous quantization of Hall conductivity at fillings $\nu = \pm (4 n +2)$ and $\pm (4 n +4)$ in weak magnetic fields, where $n=0,1,2, \cdots$\cite{novoselov, kimfirstpaper}. While the valley and the spin degrees of freedom of the chiral quasi-particles stand responsible for the four fold degeneracy of the Landau levels (LLs), the particle-hole symmetric quantization of the Hall conductivity reflects the Dirac or Dirac-like vacuum structure in these materials\cite{sharapov, falko}. Additional twofold orbital degeneracy of the zeroth LL (ZLL) in bilayer graphene (BLG) arises from the parabolic dispersion at low energies \cite{falko}, yielding a constant electronic density of states at the charge-neutrality point (CNP) in the absence of magnetic fields, which in turn enhances the effect of electron-electron interactions. Interestingly, a number of recent experiments strongly suggesting the possibility of broken-symmetry phases in BLG even without external magnetic and/or electric fields \cite{yacoby, weiss-PRL, Jairo, weiss-solstatecommun, titltedfieldBLG}. On the other hand, Dirac fermions in monolayer graphene continue to find themselves in a robust semi-metallic phase, and ordering possibly takes place only in the presence of magnetic fields. Among numerous possibilities\cite{ bitanclassification, nandkishore-levitov}, some promising candidates for the underlying ordered phases in pristine BLG are gapless \emph{nematic}\cite{kun-oskar}, and fully gapped \emph{layer antiferromagnet} (LAF) states\cite{oskar}. While the former one breaks the threefold rotational symmetry of the honeycomb lattice, the LAF order corresponds to a staggered pattern of fermion spin among the layers, which, for example, can be favored by on-site Hubbard interaction \cite{oskar, robert}.

 Splitting of the topologically protected ZLL in graphene-based systems necessarily requires the electron-electron interaction and/or Zeeman coupling of electrons spin with the magnetic field. The existence of completely filled valence band LLs, which, in principle, can get renormalized due to an ordering in the vicinity of the CNP \cite{Kveshchenko-MLG, miransky-MLG-QHE, herbut-originalQHE, bitan-oddQHE, miransky}, places the quantum Hall physics in carbon based layered materials in a different paradigm than that in regular non-relativistic  two-dimensional electron gases\cite{prange}. Therefore, the fully gapped states or ``masses", such as layer-polarized state, corresponding to an imbalance of average electronic density between two layers \cite{Mcdonald}, and LAF in BLG or N\'{e}el order in monolayer graphene \cite{herbut-originalQHE}, optimally lowers the ground state energy by mixing non-interacting electron- and hole-like LLs, and thereby pushing further down all the filled LLs, placed below the chemical potential. However, due to single-particle Zeeman coupling, the LAF order parameter(OP) in BLG gets projected onto the easy plane, in a direction perpendicular to the applied magnetic field, and a ferromagnetic order develops in its direction, resembling in this regard the situation in monolayer graphene with N\'{e}el order \cite{herbutso3}. I name this ground state easy-plane LAF(EPLAF). 

\begin{figure*}[htb]
\includegraphics[width=4.3cm,height=3.25cm]{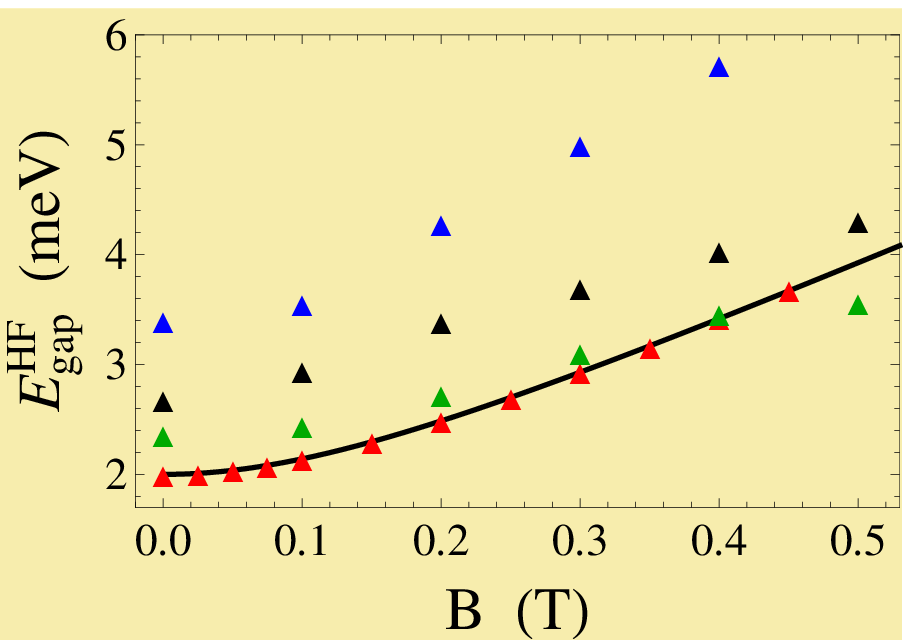}
\includegraphics[width=4.3cm,height=3.25cm]{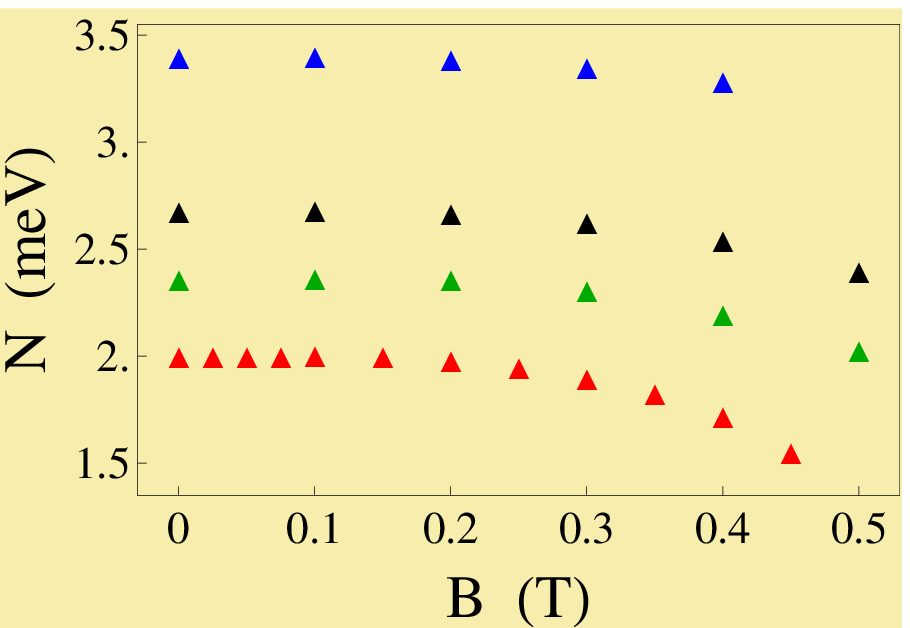}
\includegraphics[width=4.3cm,height=3.25cm]{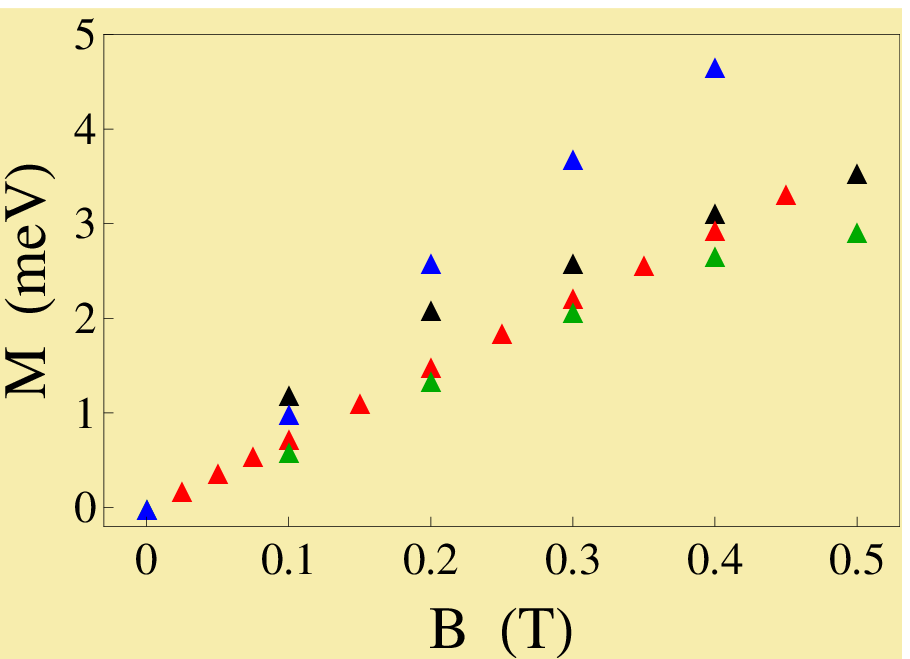}
\includegraphics[width=4.3cm,height=3.25cm]{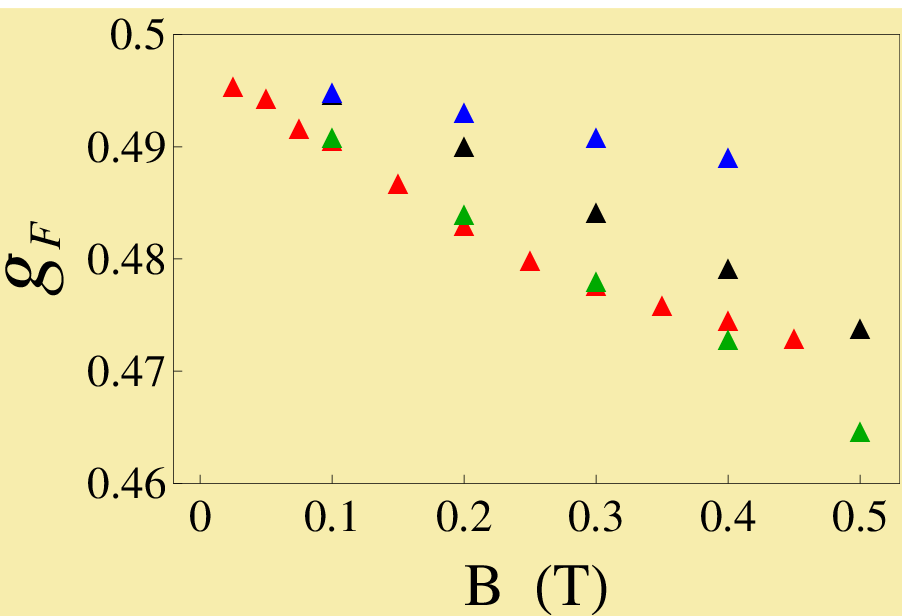}
\caption[] {(Color online) Left: Best fit to the observed gap in Ref.~\onlinecite{Jairo}(red), B2a sample in Ref.~\onlinecite{weiss-solstatecommun}(green), Ref.~\onlinecite{titltedfieldBLG} (black), B2b sample in Ref.~\onlinecite{weiss-solstatecommun} (blue) with the total gap ($E^{HF}_{gap}$), obtained self consistently \cite{supplementary}. Solid black line represents $E_{g}= \Delta + \sqrt{\Delta^2 + a^2 B^2},$ with $a=5.5$ meV/T and $\Delta \sim 1$ meV \cite{Jairo}. Second, third and forth panels respectively represent the EPLAF ($N$), FM ($M$), and ferromagnetic interaction ($g_F$), associated with the points in the leftmost panel.}\label{fitexperiment}
\end{figure*}

Although various experiments have suggested the existence of insulating BLG \cite{yacoby, weiss-PRL}, the nature of the broken symmetry phase remained puzzling for a while. Recently a well-resolved gap ($E_g$) in pristine BLG has been observed, which increases \emph{monotonically} with the magnetic field (B), conforming to a closed form $E_{g}= \Delta + \sqrt{\Delta^2 + a^2 B^2},$ where $a=5.5$ meV/T and $\Delta \sim 1$ meV \cite{Jairo}. Softening of this gap in a weak perpendicular electric field, and negligibly small two-terminal conductance ($G \sim 0.5 \mu$S $\ll 4 e^2/h$) at zero magnetic field, respectively excludes the possibility of an underlying layer-polarized state and topological quantum spin Hall insulators (QSHI)/anomalous Hall insulators in BLG. Thus far the LAF state appears to be the most promising ground state in half-filled BLG. Observation of insulating behavior in BLG has also been reported in Ref. \onlinecite{titltedfieldBLG}, and for $B2$ samples in Ref. \onlinecite{weiss-solstatecommun}. The zero magnetic field gap in these samples is $\sim 2.3-3.5$ meV, which increases linearly with the field as $\sim 3-6$meV/T. Here, I address the evolution of the LAF state in BLG under the influence of quantizing magnetic fields, and show that with the underlying EPLAF state one finds reasonably good agreement with the observed scaling of the gap in various experiments at neutral and finite fillings \cite{Jairo, weiss-solstatecommun,titltedfieldBLG, yacoby-scaling,kim-2010-nu2,velasconu2}.

The free energy in the presence of a uniform background of the electronic density, LAF ($\vec{N}$) order, and magnetization ($\vec{M}$) reads as \cite{herbutso3, herbut-book}
\begin{equation}
E_{gr}= \frac{\vec{N}^2}{4 g_A} + \frac{\vec{M}^2}{4 g_F} + E_{0} \left[ \vec{N}, \vec{M} \right].
\label{totalHF}
\end{equation}    
$E_{0} \left[ \vec{N}, \vec{M} \right]$ is the ground state energy per unit area of the effective single-particle Hamiltonian 
\begin{equation}\label{twoband1}
H_{HF}= H_0 (\lambda) - \left( \vec{N} \cdot \vec{\sigma} \right) \otimes \gamma_0 + M \left(\sigma_3 \otimes I_4\right),
\label{HFhamil}
\end{equation}    
where $H_0(\lambda)=H_0 + \lambda \left(\sigma_3 \otimes I_4\right)$, with
\begin{equation}\label{twoband2}
H_0 = \sigma_0 \otimes \left[ \gamma_2 \left( \pi^2_x-\pi^2_y \right) - \gamma_1 \left( \pi_x \pi_y+\pi_y \pi_x \right) \right]/(2 m^\ast).
\end{equation}
The effective mass of the parabolic dispersion in BLG is $m^* \approx 0.028 m_e$, where $m_e$ is the electronic mass\cite{yacoby}. The magnetic field $B=\epsilon_{3 i j} \partial_i A_j$ is set to be perpendicular to the BLG plane, so is magnetization, and $\pi_{j}=\left(-i \partial_j - A_j \right)$. The Zeeman coupling in BLG reads as $\lambda = 0.014 \omega_c$, where $\omega_c$ is the cyclotron frequency. The $\gamma$ matrices read as $\gamma_0=\sigma_0\otimes\sigma_3$, $\gamma_1=\sigma_3\otimes\sigma_2$, $\gamma_2=\sigma_0\otimes\sigma_1$, $\gamma_3=\sigma_1\otimes\sigma_2$, $\gamma_5=\sigma_2\otimes \sigma_2$, where $(\sigma_0,\vec{\sigma})$ are the two dimensional unity and Pauli matrices, respectively, and $I_4=\sigma_0 \otimes \sigma_0$ \cite{herbut-juricic-roy}.

The spectrum of $H_{HF}$ is composed of a set of LLs at well separated energies $\pm E_{n,\sigma}$, where for $\sigma=\pm 1$
\begin{equation}\label{LLenergy}
E_{n \sigma}= \left[ N^2_\perp + \left[  \sqrt{n(n-1)\omega^2_c + N^2_\parallel}  + \sigma M_T \right]^2 \right]^{1/2},
\label{energy}
\end{equation} 
with degeneracies per unit area $1/\pi l^2_B$ for $n=2,3,4, \cdots$ and $1/2 \pi l^2_B$ for $n=0,1$. Here $M_T= (\lambda + M)$ is the total magnetization, $l_B =\sqrt{\left( \hbar /e B\right)}$ is the \emph{magnetic length}, and $\vec{N}_\perp=\left( N_1,N_2 \right)$, $N_3 \equiv N_\parallel$. At half filling, LLs at negative (positive) energies are filled (empty), and therefore
\begin{equation}
E_0 \left[ \vec{N}, M \right]= - \; \frac{1}{2 \pi l^2_B} \sum_{\sigma=\pm} \big( E_{0 \sigma} + E_{1 \sigma} + 2 \sum_{n \geq 2} E_{n \sigma}  \big).
\label{HFenergy}
\end{equation}
With $|\vec{N}|$ and $N_\parallel$ as independent variables, the free energy optimization condition $\partial E_0 [ \vec{N}, M ]/\partial N_\parallel =0$ yields 
\begin{equation}
\sum_{\sigma=\pm} \big[ \sum_{n=0,1}\frac{\sigma M_T}{E_{n \sigma}} 
+ \sum_{n \geq 2} \frac{2 \sigma M_T \; N_\parallel \; (E_{n \sigma})^{-1} }{ \sqrt{N^2_\parallel + n(n-1) \omega^2_c}} \big] =0.
\end{equation} 
The left-hand side of this equation is a negative definite function of $N_\parallel$ for any nontrivial Zeeman coupling, and vanishes only for $N_\parallel \equiv 0$. Therefore, in the presence of magnetic field, LAF order gets projected onto the easy plane ($N_\parallel =0$) due to the Zeeman coupling, yielding the EPLAF state. This configuration also corresponds to the \emph{minima} of the energy. Placed in a magnetic field, an identical ground state, easy-plane N\'{e}el order, can also be realized in monolayer graphene \cite{herbutso3}.

With $N_\parallel=0$, minimizing $E_{gr}$ with respect to $M$ and $N_\perp$, we respectively obtain the coupled gap equations
\begin{eqnarray}\label{gapequaiton}
\frac{M}{g_F}=\frac{1}{4 \pi l^2_B} \sum_{\sigma=\pm} \big[ \sum_{n=0,1} \frac{M_T}{E_{n\sigma}} + \sum_{n \geq 2}\frac{\sigma \sqrt{n(n-1)} \omega_c + M_T}{E_{n\sigma}} \big] \nonumber 
\end{eqnarray}
\begin{equation}\label{gapequaitonLAF}
\frac{1}{g_A}=\frac{1}{4 \pi l^2_B} \sum_{\sigma=\pm}\big[\sum_{n=0,1} \frac{1}{E_{n\sigma}} + \sum_{n \geq 2} \frac{1}{E_{n\sigma}}\big].
\end{equation}
Within the framework of a microscopic density-density interaction, such as the on-site Hubbard model, $g_A=g_F$ at the lattice scale ($\Lambda \sim 200$meV in BLG). However, the magnetic field introduces a new length scale in the system, magnetic length $l_B$ (thus a new effective cut-off $\Lambda_B \sim (1/l_B) \ll \Lambda$), and generically $g_F=g_F(\Lambda_B) \neq g_A$. We redefine the couplings as $g_x m^\ast/(4 \pi) \rightarrow g_x$, for $x=A,F$. Besides splitting the half-filled ZLL($n=0,1$), LAF and ferromagnet (FM) OPs, respectively, pushes down and splits all the filled LLs ($n \geq 2$). As a result, the first gap equation is devoid of any divergences, while the second one exhibits an \emph{ultraviolet logarithmic divergence}, which, however, can be regularized by substituting $1=g_A\int^\Lambda_0 \left( \xi^2+\Delta^2_0\right)^{-1/2} d\xi$, ensuring the cutoff independence of the LAF OP in magnetic fields. Here $\Delta_0$ stands for the zero magnetic field LAF gap in BLG.

The same set of gap equations can also be obtained in a variation approach, developed in Ref. \cite{robert}, where the fermionic field operators are expressed as
\begin{eqnarray}
\Psi(r)=\sum_{\alpha} \left[ \psi^{(+)}_{\alpha} (r) a_{\alpha} + \psi^{(-)}_{\alpha} (r) b^\dagger_{\alpha} \right].
\end{eqnarray}          
$\psi^{(\pm)}_{\alpha}$ is the LL wave functions of $H_{HF}$ at energies $\pm E_{n\sigma}$, where $\alpha \equiv (k, n, \tau, E_{n\sigma})$, with $k$ as the wave number, and $n$, $\tau$ as the LL and valley index, respectively. $E_{n\sigma}$ is as in Eq.~(\ref{LLenergy}), but with $N_\parallel=0$ \cite{supplementary}. The variational ground state energy is $E_{V}=\langle 0 | H_{V} | 0 \rangle$, where $H_{V}$ $=H_{HF}+$ $H_{I}+$ $(H_{0}(\lambda)-H_{HF})$, and the ground state $|0\rangle$ is chosen such that $a_\alpha |0\rangle =0= b_\alpha |0\rangle$. Here $H_{I}$ is a generic four-fermion density-density interaction at the lattice scale \cite{oskar}, and the above gap equations are obtained by minimizing $E_{V}$, with respect to $M$ and $|\vec{N}_\perp|$, where $g_A=g_F=(V_0+V_{2K})$. $V_0$ and $V_{2K}$ respectively represent the forward and back-scattering interactions \cite{supplementary}.

The magnetization ($M$) increases monotonically with $g_F$, and it scales linearly with the magnetic field, when $B<0.05$T. In this regime, $N_\perp-\Delta_0 \sim B^2$, but coefficient of $B^2$ decreases with increasing $g_F$, however, very softly. For stronger magnetic fields ($B>0.1$T) $M$ scales non-linearly with $B$ for a given $g_F$, and it becomes challenging to track the scaling of $M$ with $B$ for a fixed $g_F$. Instead we search for the self-consistent solutions of $N_\perp$ and $M$, yielding reasonable agreements with the recently observed scalings of the gap at CNP \cite{Jairo, weiss-solstatecommun, titltedfieldBLG}, which here reads as $E^{HF}_{gap}=\left( N^2_\perp + (\lambda + M)^2\right)^{1/2}$. Results are shown in Fig.~1 (left), exhibiting excellent agreement with various experiments when the LAF order is accompanied by a sizable FM order (second and third panels of Fig.~1), see also Sec. IV of Ref. \onlinecite{supplementary}. Notice that quadratic scaling of the gap at low fields in Ref. \onlinecite{Jairo} crosses over to a linear one for $B \geq 0.2$T, and the scaling of $E^{HF}_{gap}$ with an underlying EPLAF state yields excellent description of these two scaling regimes. On the other hand, in Refs. \onlinecite{weiss-solstatecommun}, \onlinecite{titltedfieldBLG} gaps at the CNP have been measured for $B \geq 0.1$ T, where it scales quite linearly with $B$, and the scaling of $E^{HF}_{gap}$ is in good agreement with these observations as well.

It is interesting to note that at minimal cost of LAF order, BLG can develop a large FM order; compare second and the third panels of Fig.~1. Such peculiar behavior has a root in the fact that by depleting the LAF order system looses a significant amount of condensation energy, since the LAF order pushes down all the filled LLs below the chemical potential. The compensating FM order, which, on the other hand, lowers the ground state energy only by enhancing the splitting of the ZLL, therefore needs to be large, in agreement with the results obtained from the self-consistent calculations. The FM OP(M) scales quite linearly (third panel of Fig.~1) and the dimensionless ferromagnetic interaction ($g_F$) decreases monotonically (fourth panel of Fig.~1), with increasing $B$ along all the curves in Fig.~1(left), observed experimentally. Hence, $g_F$ exhibits {\em universal flow} towards its bare value $g^b_F=g_A$, which is $0.3<g^b_F(=g_A)<0.32$ in Refs. \onlinecite{Jairo, weiss-solstatecommun, titltedfieldBLG} as $\Lambda_B \rightarrow \Lambda$. At such strong magnetic fields, the two-band continuum description of BLG [Eqs.~(\ref{twoband2})] completely breaks down, and finite-size effects of the system become important\cite{finitesizeBLG}. However, at intermediate strength of the magnetic field BLG can be properly described by a \emph{four-band} model (including the split-off bands) in the continuum limit. Such crossover roughly takes place around $B_c \sim 2$T, when only few LLs (say $\leq 20$) are placed within the cutoff $\Lambda \sim 200$ meV for two band model \cite{cutoffcoment}. The scaling of the gap at and near the CNP in BLG beyond $B_c$ becomes qualitative similar to the one in monolayer graphene \cite{bitanscaling}, about which in a moment.    

\begin{figure}[htb]
\includegraphics[width=4.2cm,height=3.25cm]{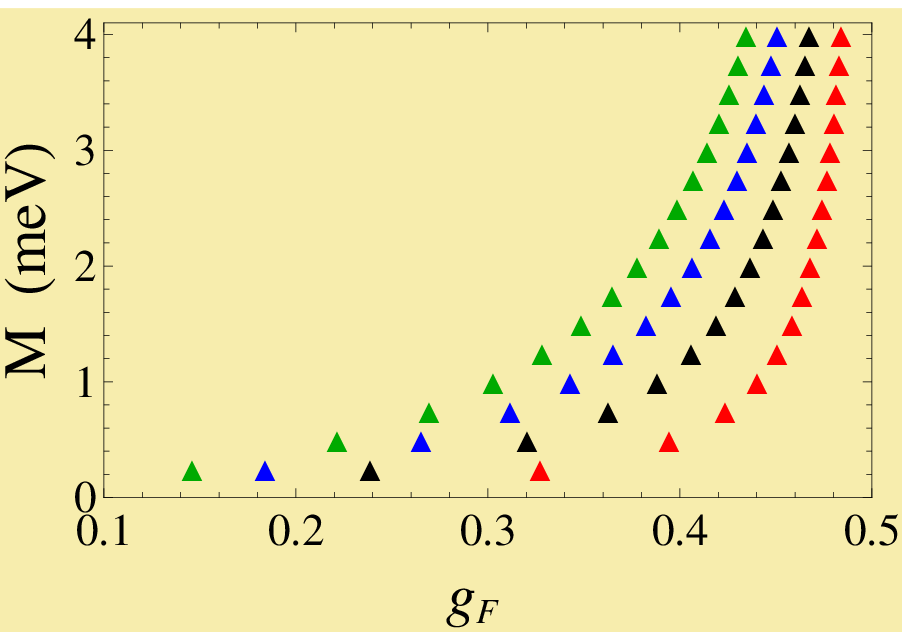}
\includegraphics[width=4.2cm,height=3.25cm]{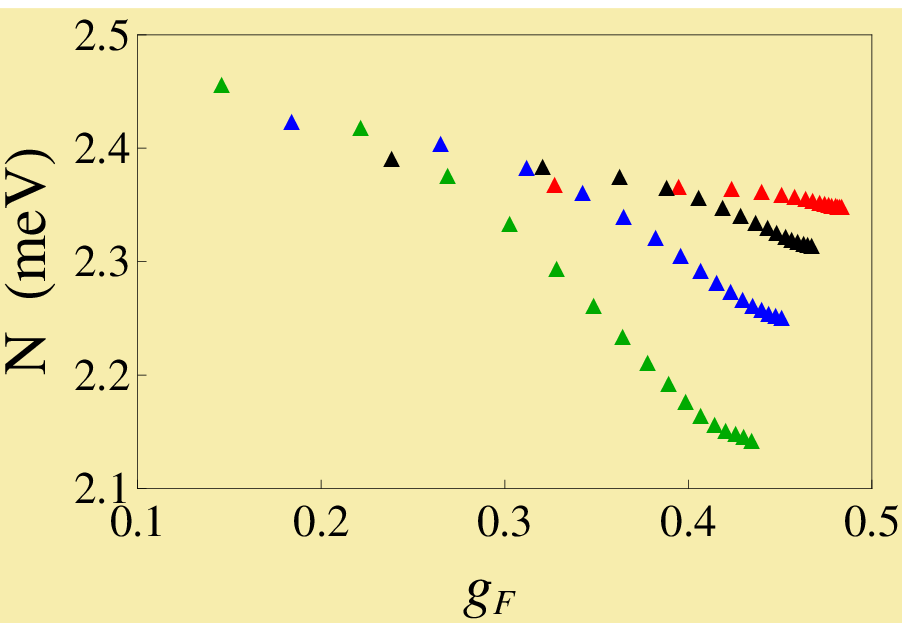}
\caption[] {(Color online) Self-consistent solution of FM (left) and EPLAF (right) OPs in tilted magnetic fields, with $B_\perp=0.1$ T (red), $0.2$ T (black), $0.3$ T (blue), $0.4$ T (green), and $B_\parallel=20 B_\perp$, as a function of $g_F$.}\label{tiltedfield}
\end{figure}

On the other hand, the excitation spectrum in the variational approach $E_{ex}=\langle E | H_{var}| E \rangle - \langle 0 | H_{var}| 0 \rangle$, where $| E \rangle = a^\dagger_{\alpha} b^\dagger_{\beta} |0\rangle$ is the excited state \cite{robert}, reads as 
\begin{equation}
E_{ex} = \left\{ 
\begin{array}{r l}
 E^{t}_{\alpha,\beta} + 2 \big( \frac{m^\ast}{4 \pi}\big) \tilde{V} \omega_c & \text{if} \; n_\alpha, n_\beta=0/1, \\ \\
E^{t}_{\alpha,\beta} + \big( \frac{m^\ast}{4 \pi}\big)  \tilde{V}  \omega_c & \text{if} \; n_\alpha= 0/1, n_\beta \geq 2 \: \text{vice-versa}, \\ \\
E^{t}_{\alpha,\beta} & \text{if} \; n_\alpha, \; n_\beta \geq 2,
\end{array} \right.
\end{equation}
where $\tilde{V}=(V_0-V_{2K})$, $E^{t}_{\alpha,\beta}=E_\alpha+E_\beta$, and $n_{\alpha/\beta}$ corresponds to the LL index of $\alpha/\beta$ \cite{supplementary}. Therefore, the excitation spectrum depends on two parameters $g_F$ and $\tilde{V}$. The interaction $\tilde{V}$ possibly captures the effect of quantum Hall ferromagnet order \cite{robert, kharitonov}. However, for $\tilde{V}=0$\cite{Hubbardmodel} the lowest energy excitation always occurs by creating particle-hole pair within the ZLL, and I obtain excellent agreements with different experiment with $E^{HF}_{gap}=E_{ex}/2$. Therefore, it appears that near the CNP Hall ferromagnet order plays a very minor role in the quantum Hall regime of BLG. However, the relative importance of these OPs in BLG can only be settled through future experiments.

It is also interesting to investigate the evolution of the EPLAF state in tilted magnetic fields. The LAF and the interaction-driven FM ($M$) OPs scale only with the perpendicular component of magnetic field ($B_\perp$), while the Zeeman term couples with the total magnetic field ($B_{t}$). Performing the same set of self-consistent calculations, however, in the presence of tilted magnetic fields, I cannot see any indication of a phase transition from EPLAF to a pure FM state, even for fixed $B_\perp=0.4$T, and a parallel component of the field $B_\parallel$ as high as $8$ T and for $0.1<g_F< 0.5$, see Fig.~2\cite{supplementary}. The existence of LAF order even without a magnetic field possibly provides such robustness to the LAF state in BLG, placed in tilted magnetic fields, which has also been demonstrated in a recent experiment \cite{titltedfieldBLG}, where the gap at CNP is found to decrease as $\sim 60\mu$eV/T $\ll \lambda$ with a \emph{perfectly} parallel magnetic field \cite{titltedfieldBLG}. Nevertheless, interactions in BLG can be weak enough, such that ordering possibly happens only in the presence of a perpendicular magnetic field, similar to what happens in monolayer graphene \cite{Kveshchenko-MLG, miransky-MLG-QHE, herbut-originalQHE, bitan-oddQHE, herbut-juricic-roy}. It is then possible to realize a transition from EPLAF to a pure FM phase, at least when $B_\perp \ll B_t$ \cite{herbutso3}. A pure FM state in BLG yields a two-terminal Hall conductance $\sigma_{xy}=4 e^2/h$ \cite{bitanclassification}, since FM and QSHI leads to identical splitting of the ZLL, which in turn supports four counter-propagating edge states ($\sigma_{xy}=2 e^2/h$ in monolayer graphene \cite{abanin, miransky-MLG-QHE-2}). On the other hand, the edge states in pure LAF/EPLAF state are fully gapped, leading to $\sigma_{xy}=0$ at the CNP. Recently, quantized two-terminal conductance of $\sigma_{xy}=4 e^2/h$ at the CNP, when $B_\perp \sim 2$T, and $B_\parallel \sim 20$ T, has been observed in a metallic BLG \cite{kim-patric}. However, quantized ($\sim 4e^2/h$) two-terminal conductance in an insulating BLG, placed in parallel magnetic field, remains to be observed.

 Placing the chemical potential close to the first excited state at $\pm E^{HF}_{gap}$, additional incompressible Hall states at filling $\nu=\pm 2$ can be formed by developing a third component of the LAF order $(N_\parallel)$, in the direction of the applied magnetic field. To the leading order in $N_\parallel$ the activation gap for $\nu=\pm 2$ Hall states reads as 
\begin{equation}
E^{(\pm 2)}_{gap}= 2 \;(\lambda+M)\;  N_\parallel /E^{HF}_{gap}+ {\cal O} (N^2_\parallel).
\label{nu2}
\end{equation}                
A similar mechanism can be responsible for the formation of $\nu=\pm 1$ Hall states in monolayer graphene \cite{herbutso3}. With an underlying EPLAF ordering at $\nu=0$, $N_\parallel$ receives contributions only from half of the ZLL, and hence $N_\parallel \sim B$, but $N_\perp > N_\parallel$. Hence, $E^{(\pm 2)}_{gap}$ is smaller than the gap for the $\nu=0$ Hall state, and that is possibly why $\nu=\pm 2$ Hall states are resolved only for $B>1$ T \cite{velasconu2}. For $B>1$T the gap at $\nu=0$ scales linearly with $B$ \cite{Jairo}, and thus the gap at $\nu=2$ should also scale linearly with $B$, in qualitative agreement with recent experimental observations\cite{velasconu2, yacoby-scaling}. Since, in the presence of perpendicular electric field $E^{HF}_{gap}$ decreases \cite{Jairo}, resultantly $E^{(\pm 2)}_{gap}$ should increase. Strong electric field induced enhancement of the gap for $\nu=2$ Hall state has already been observed experimentally \cite{velasconu2}. However, a finite $N_\parallel$ at filling $\nu=\pm 2$ also causes simultaneous layer-polarization of average electronic density. Hence, $E^{(\pm 2)}_{gap}$ can either increase or decrease with a weak electric field, depending on the relative sign of $N_\parallel$ and electric field induced layer polarizations, which may serve as a litmus test of the proposed scenario. At stronger magnetic fields, the linear scaling of the $\nu=0, \pm 2$ Hall states is expected to cross over to a $\sqrt{B}$ scaling, similar to the one in monolayer graphene \cite{bitanscaling}. Recently observed linear scaling of the gap at $\nu=0,2$ for $1$T$<B<10$T \cite{yacoby-scaling} and a $\sqrt{B}$ scaling of the $\nu=2$ Hall state for $B>10$ T \cite{kim-2010-nu2} possibly bears the signatures of such crossover scaling in BLG. Due to the enhanced interaction effect in BLG such crossover can take place within accessible range of magnetic fields.

Interaction-driven orders cannot lift the orbital degeneracy ($E_{0\sigma}\equiv E_{1\sigma}$) of the ZLL in BLG. However, the remote hopping between the sites on two layers, represented by
\begin{equation}
H=\frac{2 v_0 v_1}{t_\perp} \sigma_0 \otimes Diag.(\pi_+ \pi_-,\pi_- \pi_+,\pi_- \pi_+,\pi_+ \pi_-),
\end{equation} 
gives rise to a non interacting gap between the ZLLs with $n=0$ and $1$, since in the presence of magnetic fields, $(\pi_+ \pi_-,\pi_-\pi_+) \rightarrow 2 \hbar^2/l^2_B (1+ \hat{n},\hat{n})$, where $\hat{n}$ is the LL number operator, yielding $\nu=\pm 1$ Hall states. Here $\pi_\pm=\pi_x \pm i \pi_y$ and $v_j=t_j \sqrt{3} a/ (2 \hbar)$ for $j=0,1$. $t_0, (t_1) t_{\perp}$ are respectively the intralayer and interlayer (next-)nearest-neighbor hopping amplitudes \cite{notationhopping}. The activation gap for the $\nu=\pm 1$ Hall state, $E_{\nu=\pm 1}$ should scale linearly with the magnetic field, and with currently estimated strength for various band-parameters $E_{\nu=\pm 1} 0.2$ meV/T \cite{bandparameter}. Similar splitting can also be achieved by applying an electric field between the layers\cite{barlas}, and so far in an insulating BLG $\nu=1$ Hall plateau has only been observed in the presence of perpendicular electric fields\cite{velasconu2}. Nevertheless, in metallic BLG the $\nu=1$ Hall state has been observed at strong magnetic fields, and the gap is found to scale as $0.1$meV/T \cite{yacoby-scaling} and $0.41$ K/T\cite{kim-2010-nu2}.

Formation of fractional quantum Hall states in the ZLL depends on its degeneracy lifting at integer fillings\cite{khveshchenko}. At weak magnetic fields (and without any electric field), when the orbital degeneracy of the ZLL is protected, but $\nu=0,\pm 2$ Hall plateaus are well resolved, plateaus are expected to appear at fractional fillings $\nu=\pm 2m/(2m \pm 1) $ for $|\nu|<2$, where $m=1,2,3,\cdots$, and $\pm m/(2m \pm 1)$ is the standard Jain's sequence\cite{jain}. The additional factor of 2 in the numerator arises from the residual orbital degeneracy of the ZLL. At stronger magnetic fields, when the orbital degeneracy of the ZLL is lifted and plateaus at fillings $\nu=0,\pm 1, \pm 2$ are well resolved, BLG should discern standard Jain's sequences at fillings $\nu=\pm m/(2m \pm 1)$ and $\pm (1+m/(2m \pm 1))$. A detailed study of the fractional quantum Hall effect in BLG is quite rich, and I leave it for future investigation. Nevertheless, recently there have been suggestive signatures for the $\nu=1/3$ fractional Hall plateau in BLG, where $\nu=1$ Hall state has also been resolved \cite{laufractional}.

\emph{Acknowledgement:} I would like thank O. Vafek for suggesting this work, many useful discussions and his continued interest in this work. Author is in debt to Igor. F. Herbut for number on interesting discussions and valuable comments on this Rapid Communication. It is the author's pleasure to acknowledge fruitful discussion with C. N. Lau and J. Velasco, Jr. The author is very thankful to M. Weiss and C. Sch$\ddot{\mbox{o}}$nenberger for providing many data from Refs. \onlinecite{titltedfieldBLG}, \onlinecite{weiss-solstatecommun}. This work was supported at National High Magnetic Field Laboratory by NSF Cooperative Agreement No. DMR-0654118, the State of Florida, and the U. S. Department of Energy. I am thankful to Ecole de Physique, Les Houches for hospitality during the summer school “Strongly interacting quantum systems out of equilibrium” where a part of this work was finalized.

\vspace{14cm}

\onecolumngrid

\begin{center}
{\bf Supplementary material of ``Theory of integer quantum Hall effect in insulating bilayer graphene''}
\end{center}

\begin{center}
Bitan Roy$^{1,2}$
\end{center}
\begin{center}
$^1$ National High Magnetic Field Laboratory, Florida State University, Florida 32306, USA \\
$^2$ Condensed Matter Theory Center, Department of Physics, University of Maryland, College Park, MD 20742, USA
\end{center}
\vspace{1cm}

I here present details of the diagonalization of effective single-particle Hamiltonian $H_{HF}$, derivation of gap equations using variational approach\cite{robertsupple}, and the computation of the excitation spectrum for $\nu=0$ quantum Hall state in insulating bilayer graphene with an underlying easy-plane layer anti-ferromagnet order. Moreover, I also present some details on comparison of the gap at charge neutrality point $E^{HF}_{gap}$ in Hartree-Fock approach, with the measured gaps in various experiments\cite{LauJairo, baselSSC, baselRC}, and evolution of the layer anti-ferromagnet order in tilted magnetic fields. Let us first show some detail of the how one can arrives at the gap equations, I have presented in the main part of the paper, with only the easy-plane component of the anti-ferromagnet and the easy axis ferromagnetic order, in the presence of a magnetic field. 

\section{Variational Hamiltonian and Landau level spectrum}
The Hamiltonian describing the free motion of fermions in bilayer graphene in the presence of magnetic field reads as 
\begin{eqnarray}
H_{free}= \int d^2 r \psi^\dagger(r) \bigg\{ I_2 \otimes \gamma_2 \left( \frac{\pi^2_x-\pi^2_y}{2 m^*} \right) + I_2 \otimes \gamma_1 \left( \frac{-\pi_x \pi_y-\pi_y \pi_x}{2 m^*} \right) +\lambda \left( \sigma_3 \otimes I_4 \right) \bigg\} \psi(r)
=\int d^2 r \psi^\dagger(r) \hat{H}_{free}\psi(r),
\end{eqnarray}
$\lambda$ is the single particle Zeeman coupling of electrons spin with the magnetic field, set perpendicular to the bilayer graphene plane. The orbital effect of the magnetic field is captured via minimal substitution $\pi_{j}=\left(-i \partial_j - A_j \right)$, with $j=x,y$ and strength of the magnetic field reads as $B=\epsilon_{3 i j} \partial_i A_j$. The eight component fermionic field is defined as $\psi=\left[ \psi_+, \psi_- \right]^\top$, where 
\begin{equation}
\psi^\top_\sigma =\left[ v_{1, \sigma} (\vec{K}+\vec{q}), v_{2, \sigma} (\vec{K}+\vec{q}), v_{1, \sigma} (-\vec{K}+\vec{q}), v_{2, \sigma} (-\vec{K}+\vec{q}) \right],
\end{equation} 
and $\sigma=\pm$ are the projections of electrons spin along the z-direction. This representation is spin rotationally invariant and therefore our formalism can be extended easily even when the field is tilted. A generic four fermion interactions in bilayer graphene is described by the interacting Hamiltonian\cite{vafekRGpaper}
\begin{eqnarray}
H^{(4)}_{int}&=&\sum^2_{j=1}\int  d^2r \bigg\{ \frac{V_0}{2A_{uc}} \left[\left(\psi^{\dagger}\mathcal{M}^{(f)}_j\psi\right)
\left(\psi^{\dagger}\mathcal{M}^{(f)}_j\psi\right)\right]
+ \frac{V_{2K}}{2A_{uc}} \bigg[ \left(\psi^{\dagger}\mathcal{M}^{(b)}_j\psi\right)
\left(\psi^{\dagger}{\mathcal{M}^{(b)}_j}^{T}\psi\right) \nonumber  \\
&+& \left(\psi^{\dagger}{\mathcal{M}^{(b)}_j}^T\psi\right) 
\left(\psi^{\dagger}\mathcal{M}^{(b)}_j\psi\right)\bigg] \bigg\}. 
\end{eqnarray}
$V_0$ and $V_{2K}$ respectively corresponds to the strength of forward and back scattering interactions. Onsite Hubbard model is also described by $H^{(4)}_{int}$, with a constraint $V_0=V_{2K}$.\cite{vafekRGpaper} Various matrices appearing in $H^{(4)}_{int}$ are defined as 
\begin{eqnarray}
\mathcal{M}^{(f)}_1\!\!=\!\!I_2 \otimes \left(\begin{array}{cccc}
1 & 0 & 0 &0 \\
0 & 0 & 0 &0 \\
0 & 0 & 1 &0 \\
0 & 0 & 0 &0
\end{array}\right),\;
\mathcal{M}^{(f)}_2\!\!=\!\!I_2 \otimes \left(\begin{array}{cccc}
0 & 0 & 0 &0 \\
0 & 1 & 0 &0 \\
0 & 0 & 0 &0 \\
0 & 0 & 0 &1
\end{array}\right),
\mathcal{M}^{(b)}_{1}\!\!=\!\! I_2 \otimes \left(\begin{array}{cccc}
0 & 0 & 1 &0 \\
0 & 0 & 0 &0 \\
0 & 0 & 0 &0 \\
0 & 0 & 0 &0
\end{array}\right),\;
\mathcal{M}^{(b)}_{2}\!\!=\!\! I_2 \otimes \left(\begin{array}{cccc}
0 & 0 & 0 &0 \\
0 & 0 & 0 &1 \\
0 & 0 & 0 &0 \\
0 & 0 & 0 &0
\end{array}\right).
\end{eqnarray}

To perform the variational mean field calculation, I add and subtract the layer anti-ferromagnet (LAF) and the ferromagnet (FM) order parameters (source terms),
\begin{equation}\label{OPHamil}
H_{OP}=\vec{N} \cdot \int d^2 r \: \psi^\dagger(r) \left[ \vec{\sigma} \otimes \gamma_0 \right] \psi(r) \quad  
+ M \int d^2 r \: \psi^\dagger(r) \left[ \sigma_3 \otimes I_4 \right] \psi(r)=\int d^2 r \: \psi^\dagger(r) \hat{H}_{OP} \psi(r).
\end{equation}
For now, I keep the orientation of the anti-ferromagnet order parameter ($\vec{N}$) arbitrary, but restrict the ferromagnet order parameter ($M$) only along the applied magnetic field. Next I compute the energy spectrum of the auxiliary Hamiltonian 
\begin{equation}
\hat{H}_{aux}= \hat{H}_{free} + \hat{H}_{OP} \equiv H_{HF}.
\end{equation}
To diagonalize the auxiliary Hamiltonian, $H_{aux}$, it is worth noticing that two valleys remain decoupled, even in the presence of layer anti-ferromagnet and ferromagnet orders. One can therefore, bring $H_{aux}$ in block diagonal. It can be achieved by exchanging the 2nd and the 3rd $2 \times 2$ block of $H_{aux}$, yielding $H_{aux} \rightarrow H_+ \oplus H_-$, where
\begin{eqnarray}
H_\pm= I_2 \otimes \sigma_1 \left( \frac{\pi^2_x-\pi^2_y}{2 m^*} \right) \pm I_2 \otimes \sigma_2 \left( \frac{-\pi_x \pi_y-\pi_y \pi_x}{2 m^*} \right)
+ \vec{N} \cdot \left( \vec{\sigma} \otimes \sigma_3 \right)+ \left( \lambda+M \right) \sigma_3 \otimes I_2.
\end{eqnarray} 
However, both $H_\pm$ are unitarily equivalent to a generic Hamiltonian,
\begin{equation}
H=\gamma_2 \left( \frac{\pi^2_x-\pi^2_y}{2 m^*} \right) + \gamma_1 \left( \frac{-\pi_x \pi_y-\pi_y \pi_x}{2 m^*} \right) -N_1 \gamma_3 -N_2 \gamma_5 -N_3 \gamma_{0} + \left( \lambda+M\right) \gamma_{35}.
\end{equation}
Explicitly, $H_1 = U^\dagger_1 H U_1$ where $U_1= I_2 \oplus (-i \sigma_1)$ and $H_2 = U^\dagger_2 H U_2$ with $U_2= (-i \sigma_1) \oplus I_2$, but $N_3 \rightarrow -N_3$. In a similar way one can also diagonalize the effective single-particle Hamiltonian for single layer graphene when a N\'{e}el order develops at the charge-neutrality point in the presence of magnetic fields, originally shown in Ref.~\onlinecite{herbutso3AF}. However, the structure of two matrices $U_1$ and $U_2$ are slightly different for monolayer and bilayer graphene. Energy spectrum can then be immediately computed yielding a set of Landau levels at $\pm E_{n,\sigma}$, where 
\begin{equation}\label{spectrum}
E_{n,\sigma}=\left[  |N_\perp|^2 + \left( \left[ n(n-1)\omega^2_c + N^2_3 \right]^{1/2} + \sigma \left( \lambda + M \right) \right)^2 \right]^{1/2},
\end{equation}
with degeneracy per unit area $1/2 \pi l^2_B$ for $n \geq 2$, and $1/\pi l^2_B$ for $n=0,1$, where $l_B$ is the magnetic length, $|N_\perp|=\sqrt{N^2_1+N^2_2}$ and $\omega_c$ is the cyclotron frequency. Next I wish to find the orientation of the anti-ferromagnet order that minimizes the ground state energy of filled Fermi sea. At half-filling all the states at negative energies are completely filled, while those at positive energies are completely empty. Therefore, the Hartree-Fock ground state energy of the single particle auxiliary Hamiltonian reads as 
\begin{equation}
E_0 \left[ \vec{N}, M \right]= - \; \frac{1}{2 \pi l^2_B} \sum_{\sigma=\pm} \left( E_{0 \sigma} + E_{1 \sigma} + 2 \sum_{n \geq 2} E_{n \sigma}  \right).
\end{equation}
To find the configuration of $\vec{N}$ to minimize the Hartree-Fock ground state energy, I choose $|\vec{N}_{\perp}|$ and $N_3(=N_\parallel)$ as independent variables. Then the energy minimization condition 
\begin{equation}
\frac{\partial E_0 \left[ \vec{N}, M \right]}{\partial N_3} =0 \: \Rightarrow \:
\sum_{\sigma=\pm} \sigma \left( \lambda + M\right) \left[ \sum_{n=0,1}\frac{1}{E_{n \sigma}} 
+ \sum_{n \geq 2} \frac{2 N_3}{E_{n \sigma} \sqrt{N^2_3 + n(n-1) \omega^2_c}} \right] = 0.
\end{equation} 
The left hand side of this equation is a negative definite function of $N_3$ and vanishes only for $N_3 \equiv 0$. Therefore, in the presence of the Zeeman coupling the anti-ferromagnet order is projected in a plane perpendicular to the direction of the magnetic field, the spin-easy-plane. One can as well check that such configuration corresponds to the minima of the energy. From now I, therefore set $N_3 \equiv 0$, and $|N_\perp| \equiv N$, for notational simplicity.

The Landau level wave functions for $n=0,1$, however localized near $+ \vec{K}$ valley at energy $E_0=\sqrt{N^2+(\lambda+M)^2}$ reads as
\begin{equation}
\bigg| + E_0 \bigg\rangle =  
\left[ \begin{array}{c}
b_0 \chi_n \\
0 \\
a_0 \chi_n \\
0
\end{array}\right], \quad 
\bigg| - E_0 \bigg\rangle =  
\left[ \begin{array}{c}
-a_0 \chi_n \\
0 \\
b_0 \chi_n \\
0
\end{array}\right],
\end{equation}
while those residing near the valley at $-\vec{K}$ assume the form
\begin{equation}
\bigg| + E_0 \bigg\rangle =  
\left[ \begin{array}{c}
0 \\
- b_0 \chi_n \\
0 \\
a_0 \chi_n 
\end{array}\right], \quad 
\bigg| - E_0 \bigg\rangle =  
\left[ \begin{array}{c}
0 \\
a_0 \chi_n \\
0 \\
b_0 \chi_n \\
\end{array}\right].
\end{equation}
With $M_T=\lambda+M$ as total magnetization, I have 
\begin{equation}
a_0=\frac{1}{\sqrt{2}} \sqrt{1-\frac{M}{E_0}}, \quad 
b_0=\frac{1}{\sqrt{2}} \sqrt{1+\frac{M}{E_0}}. 
\end{equation}
The wave-function of the Landau levels at energies $E_{n,\sigma}$ for $n \geq 2$, localized in the vicinity of $+\vec{K}$ valley are 
\begin{equation}
\bigg| + E_{n,-} \bigg\rangle =  
\left[ \begin{array}{c}
a_n \chi_n \\
- a_n \chi_{n-2} \\
b_n \chi_n \\
b_n \chi_{n-2} 
\end{array}\right], \: \:
\bigg| - E_{n,-} \bigg\rangle =  
\left[ \begin{array}{c}
- b_n \chi_n \\
b_n \chi_{n-2} \\
a_n \chi_n \\
a_n \chi_{n-2} \\
\end{array}\right], \: \:
\bigg| + E_{n,+} \bigg\rangle =  
\left[ \begin{array}{c}
-c_n \chi_n \\
- c_n \chi_{n-2} \\
-d_n \chi_n \\
d_n \chi_{n-2} 
\end{array}\right], \: \: 
\bigg| - E_{n,+} \bigg\rangle =  
\left[ \begin{array}{c}
d_n \chi_n \\
d_n \chi_{n-2} \\
-c_n \chi_n \\
c_n \chi_{n-2} \\
\end{array}\right],
\end{equation}
where
\begin{equation}
a_n=\frac{1}{2} \: \sqrt{1-\frac{e_n-M_T}{E_{n,-}}}, \quad
b_n=\frac{1}{2} \: \sqrt{1+\frac{e_n-M_T}{E_{n,-}}}, \quad
c_n=\frac{1}{2} \: \sqrt{1+\frac{e_n+M_T}{E_{n,+}}}, \quad 
d_n=\frac{1}{2} \: \sqrt{1-\frac{e_n+M_T}{E_{n,+}}},
\end{equation}
and $e_n=\omega_c \sqrt{n(n-1)}$. The wave functions of the Landau levels at $E_{n,\sigma}$ for $n \geq 2$ in the vicinity of $\pm \vec{K}$ are otherwise identical.
 
\section{Variational Hartree-Fock energy and gap equations}

Next I evaluate the variation ground state energy of the total Hamiltonian, $H_{aux}+H^{(4)}_{int}+H^{(2)}_{int}$ where $H^{(2)}_{int}=- H_{OP}$, in the presence of quantizing magnetic field, which quenches the spectrum of the quasi-particle into a set of Landau levels, obtained from the diagonalization of the auxiliary Hamiltonian $H_{aux}\equiv H_{HF}$. It is then worth to rewrite the fermionic field as 
\begin{equation}
\psi(r)=\sum_{k, n, \tau, E_{n\sigma}} \left[ \psi_{k, n, \tau, E_{n\sigma}} (r) a_{k, n, \tau, E_{n\sigma}} + \psi_{k, n, \tau, E_{n\sigma}} (r) b^\dagger_{k, n, \tau, E_{n\sigma}} \right],
\end{equation}
where $k$ is the wavenumber, $n$ is the Landau level index, $\tau$ is the valley index, and $E_{n\sigma}$ is the energies of the Landau levels with finite Zeeman coupling and $\sigma=\pm$ shown in Eq.~(\ref{spectrum}), after setting $N_3=N_\parallel=0$. In the above expression the term with annihilation operator $a$ gets summed over all the empty states at positive energies, while the other one with creation operator $b^\dagger$ gets summed over all the filled states at negative energies. In the presence of the Zeeman coupling and layer anti-ferromagnet order, spin is no longer a good quantum number. Hence, instead of electrons spin, I identify a new \emph{effective} quantum number $E_{n\sigma}$, the energies of the Landau levels of $H_{aux}$ to complete the Landau level basis.

The auxiliary Hamiltonian, $H_{aux}$ in the terms of the Landau level creation and annihilation operators reads as 
\begin{equation}
H_{aux}=\sum_{k,n,\tau, E_{n,\sigma}} \bigg( |E_{n,\sigma}| a^\dagger_{k, n, \tau, E_{n,\sigma}} a_{k, n, \tau, E_{n,\sigma}} + |E_{n,\sigma}| b^\dagger_{k, n, \tau, E_{n,\sigma}} b_{k, n, \tau, E_{n,\sigma}} - |E_{n,\sigma}| \bigg)
\end{equation}   
The ground state $| 0\rangle$ is chosen such that both $a$ and $b$ annihilates $| 0\rangle$. The ground state expectation value of $H_{aux}$ is
\begin{equation}
\langle 0 | H_{aux} |0 \rangle=- 2 D \sum_{n \geq 2,\sigma} E_{n,\sigma} -2 \times 2 D \; E_{0}, 
\end{equation}
where $D=1/2 \pi l^2_B$.

Next I compute the ground state energy of the interacting part of the total Hamiltonian. Let us start our discussion with the quadratic piece of the interacting Hamiltonian $H^{(2)}_{int}$. The ground state expectation value of $H^{(2)}_{int} (=-H_{OP})$ reads as 
\begin{eqnarray}
\langle 0| H^{(2)}_{int} | 0 \rangle = - \bigg\{ M \bigg[ e_{n} \bigg( \frac{1}{E_{n,+}}-\frac{1}{E_{n,-}}\bigg) + M_T \bigg( \frac{1}{E_{n,+}}+\frac{1}{E_{n,-}}\bigg)+ \frac{2 M_T}{E_{0}}\bigg] 
+ N^2 \bigg[ \frac{1}{E_{n,+}} + \frac{1}{E_{n,-}} + \frac{2}{E_{0}} \bigg] \bigg\} \times \; 2 D.
\end{eqnarray}
Finally I compute the ground state expectation value of the quartic interaction $H^{(4)}_{int}$, which reads as 
\begin{equation}
\langle 0 | \left[\psi^\dagger(r) {\cal O}_1 \psi(r) \right] \left[\psi^\dagger(r) {\cal O}_2 \psi(r) \right] | 0 \rangle =
\sum_{n<0,p>0} \left[\psi^\dagger_{n}(r) {\cal O}_1 \psi_{p} (r) \right] \left[ \psi^\dagger_p (r) {\cal O}_2 \psi_n (r) \right] + 
\sum_{n,p<0} \left[\psi^\dagger_{n}(r) {\cal O}_1 \psi_n (r) \right]  \left[ \psi^\dagger_p (r) {\cal O}_2 \psi_p (r) \right],  
\end{equation}
where $n$ and $p$ are Landau level indices. After a long and tedious calculation, one can compactly write the ground state expectation value of the quartic interaction terms as 
\begin{eqnarray}
&&\frac{\langle 0 | H^{(4)}_{int} | 0 \rangle}{D^2} = V_{2 K} \sum_{n,p,\sigma=\pm} \bigg\{ -N^2 \bigg[ \frac{1}{E_{n,+}} + \frac{1}{E_{n,-}}\bigg] \bigg[\frac{1}{4} \bigg( \frac{1}{E_{p,+}} + \frac{1}{E_{p,-}} \bigg) + \frac{1}{E_0}\bigg] + \bigg[ 2+ \sigma \frac{e_n- M_T}{E_{n,-}}- \sigma \frac{e_n+ M_T}{E_{n,+}}\bigg] \nonumber \\
&\times&\bigg[ \frac{1}{4} \bigg(2- \sigma \frac{e_p- M_T}{E_{p,-}} + \sigma \frac{e_p+ M_T}{E_{p,+}} \bigg) + \bigg(1+ \sigma \frac{M_T}{E_0} \bigg) \bigg] \bigg\} + 4 V_0 \sum_{n,p} \bigg\{ \bigg( 1+ \frac{M_T}{E_0}\bigg) +\frac{1}{4} \bigg(2- \frac{e_p-M_T}{E_{p,-}} +  \frac{e_p+M_T}{E_{p,+}}\bigg) \bigg\} \nonumber \\
&\times&   \bigg\{ \bigg( 1- \frac{M_T}{E_0}\bigg) +\frac{1}{4} \bigg(2+ \frac{e_n-M_T}{E_{n,-}} 
-  \frac{e_n+M_T}{E_{n,+}}\bigg) \bigg\} - N^2 \bigg[\frac{1}{E_0}+\frac{1}{4} \bigg(\frac{1}{E_{n,+}} +\frac{1}{E_{n,-}}\bigg) \bigg] \; \bigg[\frac{1}{E_0}+\frac{1}{4} \bigg(\frac{1}{E_{p,+}} +\frac{1}{E_{p,-}}\bigg) \bigg] \nonumber \\
&+&V_0 \sum_{n,p} \bigg\{ -\frac{N^2}{4} \bigg[\frac{1}{E_{n,+}}+ \frac{1}{E_{n,-}}\bigg] \bigg[\frac{1}{E_{p,+}}+ \frac{1}{E_{p,-}}\bigg] + \frac{1}{4} \bigg[ 2+ \frac{e_n-M_T}{E_{n,-}} -\frac{e_n+M_T}{E_{n,+}} \bigg] \; \bigg[ 2+ \frac{e_p-M_T}{E_{p,-}} -\frac{e_p+M_T}{E_{p,+}} \bigg].
\end{eqnarray}
Next I minimize the total variation energy $E_{var}= \langle 0 | H_{aux} + H^{(4)}_{int}  +  H^{(2)}_{int} | 0 \rangle $ with respect to the ferromagnet ($M$)  and the easy plane layer anti-ferromagnet order ($N$) order parameters to arrive at the gap equations, which are 
\begin{eqnarray}
M-\bigg(\frac{V_0+V_{2K}}{4 \pi l^2_B}\bigg) \bigg[ \sum_{n \geq 2} \bigg( \frac{e_n+ M_T}{E_{n,+}} + \frac{e_n - M_T}{E_{n,-}} \bigg) +   \frac{2 M_T}{E_0} \bigg]&=&0, \\
1- \bigg(\frac{V_0+V_{2K}}{4 \pi l^2_B}\bigg)  \bigg[\sum_{n \geq 2} \bigg( \frac{1}{E_{n,+}} + \frac{1}{E_{n,-}} \bigg) + \frac{2}{E_0}\bigg]&=&0,
\end{eqnarray}
as shown in the main part of the paper, with $V_0+V_{2K}=g_F$ in the first gap equation, and $V_0+V_{2K}=g_A$ in the second one. 

\section{Excitation spectrum for $\nu=0$ quantum Hall state}

I next compare the the single excitation gap of the auxiliary Hamiltonian $H_{aux}$, namely $E^{HF}_{gap}=\sqrt{N^2+ \left( \lambda+ M\right)^2}$, with the one I obtain from variational approach. To extract the excitation gap in this approach let us construct an excited state $| E\rangle$ out of the ground state $|0 \rangle$, as $| E \rangle= a^\dagger_\alpha b^\dagger_\beta |0\rangle$, where $\alpha$, $\beta$ contains all the four indices to complete the basis, and for example $ \alpha \equiv\left( k, \alpha, \tau, E_{{\alpha},\sigma} \right)$. 

Let us first compute the expectation value of the non-interacting piece of the Hamiltonian the excitation energy $\langle E| H_{aux}|E\rangle$ in the excited state. A straight forward computation yields 
\begin{equation}
\langle E| H_{aux}|E\rangle = E_\alpha + E_\beta- \sum_{n,\sigma} E_{n,\sigma} 
\end{equation}    
Next I evaluate the expectation value of the quadratic piece of the interacting Hamiltonian in the excited state $\langle E| H_{OP}|E\rangle$. This quantity depends on the index $\alpha$ and $\beta$, and is given by a simple expression 
\begin{equation}
\langle E| H_{OP}|E\rangle= -\Delta_{FM} \bigg\{ \Psi^\dagger_\alpha \left(\sigma_3 \otimes I_4 \right) \Psi_\alpha +  \Psi^\dagger_\beta \left(\sigma_3 \otimes I_4 \right) \Psi_\beta\bigg\}-\Delta_{AF} \bigg\{ \Psi^\dagger_\alpha \left(\vec{\sigma}_\perp \otimes \gamma_0 \right) \Psi_\alpha +  \Psi^\dagger_\beta \left(\vec{\sigma}_\perp \otimes \gamma_0 \right) \Psi_\beta\bigg\},
\end{equation}
where $\Psi_\alpha$ and $\Psi_\beta$ are the respective the Landau level wave functions with positive and negative energies, shown in Sec. I. 

Next I compute the expectationvalue of the four-fermion interactions in the excited state. After completing some standard algebra, I obtain
$$\langle E | \left[\psi^\dagger(r) {\cal O}_1 \psi(r) \right] \left[\psi^\dagger(r) {\cal O}_2 \psi(r) \right] | E \rangle -
\langle 0 | \left[\psi^\dagger(r) {\cal O}_1 \psi(r) \right] \left[\psi^\dagger(r) {\cal O}_2 \psi(r) \right] | 0 \rangle$$
\begin{eqnarray}
&=&\sum_{\eta>0} \left[ \Psi^\dagger_\alpha {\cal O}_1 \Psi_\eta \right] \; \left[ \Psi^\dagger_\eta {\cal O}_2 \Psi_\alpha \right]
-\sum_{\eta>0} \left[ \Psi^\dagger_\beta {\cal O}_1 \Psi_\eta \right] \; \left[ \Psi^\dagger_\eta {\cal O}_2 \Psi_\beta \right]
-\sum_{\eta<0} \left[ \Psi^\dagger_\eta {\cal O}_1 \Psi_\alpha \right] \; \left[ \Psi^\dagger_\alpha {\cal O}_2 \Psi_\eta \right]
+\sum_{\eta<0} \left[ \Psi^\dagger_\eta {\cal O}_1 \Psi_\beta \right] \; \left[ \Psi^\dagger_\beta {\cal O}_2 \Psi_\eta \right] \nonumber \\
&+& \left[ \Psi^\dagger_\alpha {\cal O}_1 \Psi_\alpha \right] \; \sum_{\eta<0} \left[ \Psi^\dagger_\eta {\cal O}_2 \Psi_\eta \right]
+\left[ \Psi^\dagger_\alpha {\cal O}_2 \Psi_\alpha \right] \; \sum_{\eta<0} \left[ \Psi^\dagger_\eta {\cal O}_1 \Psi_\eta \right]
-\left[ \Psi^\dagger_\beta {\cal O}_1 \Psi_\beta \right] \; \sum_{\eta<0} \left[ \Psi^\dagger_\eta {\cal O}_2 \Psi_\eta \right]
-\left[ \Psi^\dagger_\beta {\cal O}_2 \Psi_\beta \right] \; \sum_{\eta<0} \left[ \Psi^\dagger_\eta {\cal O}_1 \Psi_\eta \right] \nonumber \\
&-& \left[ \Psi^\dagger_\alpha {\cal O}_1 \Psi_\alpha \right] \; \left[ \Psi^\dagger_\beta {\cal O}_2 \Psi_\beta \right]
-\left[ \Psi^\dagger_\alpha {\cal O}_2 \Psi_\alpha \right] \; \left[ \Psi^\dagger_\beta {\cal O}_1 \Psi_\beta \right]
+ \left[ \Psi^\dagger_\beta {\cal O}_1 \Psi_\alpha \right] \; \left[ \Psi^\dagger_\alpha {\cal O}_2 \Psi_\beta \right]
+ \left[ \Psi^\dagger_\alpha {\cal O}_1 \Psi_\beta \right] \; \left[ \Psi^\dagger_\beta {\cal O}_2 \Psi_\alpha \right].
\end{eqnarray}

After a lengthy and tedious computation, I obtain the expression for the many-body excitation gap for the $\nu=0$ Hall state as
\begin{equation}
E_{excitation} = \left\{ 
\begin{array}{r l}
E_{\alpha} + E_{\beta} + 2 \; \big( \frac{m^\ast}{4 \pi}\big) \; \left( V_0 -V_{2K}\right) \; \omega_c & \text{if} \: n_\alpha, n_\beta=0, \text{or} \; 1, \\ \\
E_{\alpha} + E_{\beta} + \big( \frac{m^\ast}{4 \pi}\big) \; \left( V_0 -V_{2K}\right) \; \omega_c & \text{if} \: n_\alpha= 0 \; \text{or} \; 1, \; n_\beta \geq 2; \: \: \text{vice-versa}, \\ \\
E_{\alpha} + E_{\beta} & \text{if} \: n_\alpha, \; n_\beta \geq 2
\end{array} \right.
\end{equation}
as shown in the main part of the paper. The term proportional to $(V_0-V_{2K})$ possibly captures the effect of quantum Hall ferromagnet. However, excellent agreement with number of experiments is achieved for $V_0-V_{2K}=0$, when the layer-anti ferromagnet order is accompanied by a sizable interaction induced ferromagnetic order ($M$). Also notice that within the framework of Hubbard model at the lattice scale $V_0=V_{2K}$. 

\section{Comparison with experiments}

I here provide the detail comparison of self-consistently calculated gap for the $\nu=0$ Hall state with various experiments\cite{LauJairo, baselSSC, baselRC}. In my numerical calculation I do not evaluate the gap at zero magnetic field, rather use it as an input to calculate the gap at finite magnetic fields. I am extremely thankful and in debt to Markus Weiss and Cristian Sch$\ddot{\mbox{o}}$nenberger for providing number of data from their papers\cite{baselSSC, baselRC}.
\begin{table}[h]
  \begin{tabular}{|c||c|c|c|}
    \hline
   $B$(Tesla)  & {Gap in Ref.\onlinecite{LauJairo}}(meV) & {$E^{HF}_{gap}$ (meV)} & {Gap with pure LAF order}\\
   \hline \hline
    {$0.000$} & $2.00000$ & $2.00000$ & $2.00000$  \\
   \hline
    {$0.025$} & $2.00941$ & $2.00977$ & $2.00072$  \\
   \hline
    {$0.050$} & $2.03712$ & $2.03835$ & $2.00289$  \\
   \hline
    {$0.075$} & $2.08174$ & $2.08167$ & $2.00649$  \\
   \hline
    {$0.100$} & $2.14127$ & $2.13877$ & $2.01154$   \\
   \hline
    {$0.150$} & $2.29639$ & $2.29567$ & $2.02593$   \\
   \hline
    {$0.200$} & $2.48661$ & $2.49044$ & $2.04601$   \\
   \hline
    {$0.250$} & $2.70018$ & $2.69976$ & $2.07164$   \\
   \hline
    {$0.300$} & $2.92938$ & $2.93049$ & $2.10252$    \\
   \hline
    {$0.350$} & $3.16925$ & $3.16552$ & $2.13820$   \\
   \hline
    {$0.400$} & $3.41661$ & $3.42187$ & $2.17812$    \\
   \hline
    {$0.450$} & $3.66939$ & $3.67697$ & $2.22168$  \\
       \hline
  \end{tabular}
  \caption{I here extract the value of the observed gap in Ref.\onlinecite{LauJairo} substituting $B$ in the fitting function proposed in Ref.\onlinecite{LauJairo} $E_g=\Delta+\sqrt{a^2 B^2 + \Delta^2}$, with $\Delta=1$ meV, and $a=5.5$ meV/T.}
  \end{table}

\begin{table}[h]
  \begin{tabular}{|c||c|c|c|}
    \hline
   $B$(Tesla)  & {Gap in B2a sample of Ref.\onlinecite{baselSSC}}(meV) & {$E^{HF}_{gap}$ (meV)} & {Gap with pure LAF order}\\
   \hline \hline
   {$0.000$} & $2.36 \pm 0.1$ & $2.36000$ & $2.36000$  \\
   \hline
   {$0.100$} & $2.44 \pm 0.1$ & $2.44263$  & $2.36978$ \\
   \hline
   {$0.200$} & $2.72 \pm 0.1$ & $2.72276$ & $2.39904$  \\
   \hline
   {$0.300$} & $3.10 \pm 0.1$ & $3.10771$ & $2.44739$ \\
   \hline
   {$0.400$} & $3.46 \pm 0.1$ & $3.46691$ & $2.51335$  \\
   \hline
   {$0.500$} & $3.56 \pm 0.1$ & $3.56602$ & $2.59408$ \\
\hline
   \end{tabular}
   \end{table}

\begin{table}[h]
  \begin{tabular}{|c||c|c|c|}
    \hline
   $B$(Tesla)  & {Gap in B2b sample of Ref.\onlinecite{baselSSC}}(meV) & {$E^{HF}_{gap}$ (meV)} & {Gap with pure LAF order}\\
   \hline \hline
   {$0.000$} & $3.40$ & $3.40000$ & $3.40000$  \\
   \hline
   {$0.100$} & $3.55$ & $3.54891$ & $3.40679$  \\
   \hline
   {$0.200$} & $4.28$ & $4.27553$ & $3.42714$  \\
   \hline
   {$0.300$} & $5.00$ & $4.99615$ & $3.46096$  \\
   \hline
   {$0.400$} & $5.73$ & $5.71896$ & $3.50802$  \\
\hline
   \end{tabular}
   \end{table}

\begin{table}[h]
  \begin{tabular}{|c||c|c|c|}
    \hline
   $B$(Tesla)  & {Gap in Ref.\onlinecite{baselRC}}(meV) & {$E^{HF}_{gap}$ (meV)} & {Gap with pure LAF order}\\
   \hline \hline
   {$0.000$} & $2.67 (\pm 1\%)$ & $2.67683$ & $2.67683$  \\
   \hline
   {$0.100$} & $2.94 (\pm 1\%)$ & $2.93987$ & $2.68546$  \\
   \hline
   {$0.200$} & $3.39 (\pm 1\%)$ & $3.39727$ & $2.71127$  \\
   \hline
   {$0.300$} & $3.69 (\pm 1\%)$ & $3.69952$ & $2.75407$  \\
   \hline
   {$0.400$} & $4.03 (\pm 1\%)$ & $4.03338$ & $2.81303$  \\
   \hline
   {$0.500$} & $4.30 (\pm 1\%)$ & $4.29221$ & $2.88634$  \\
\hline
   \end{tabular}
   \end{table}

\pagebreak

\section{Additional numerical results for tilted magnetic fields}

\begin{figure*}[htb]
\includegraphics[width=4.3cm,height=3.25cm]{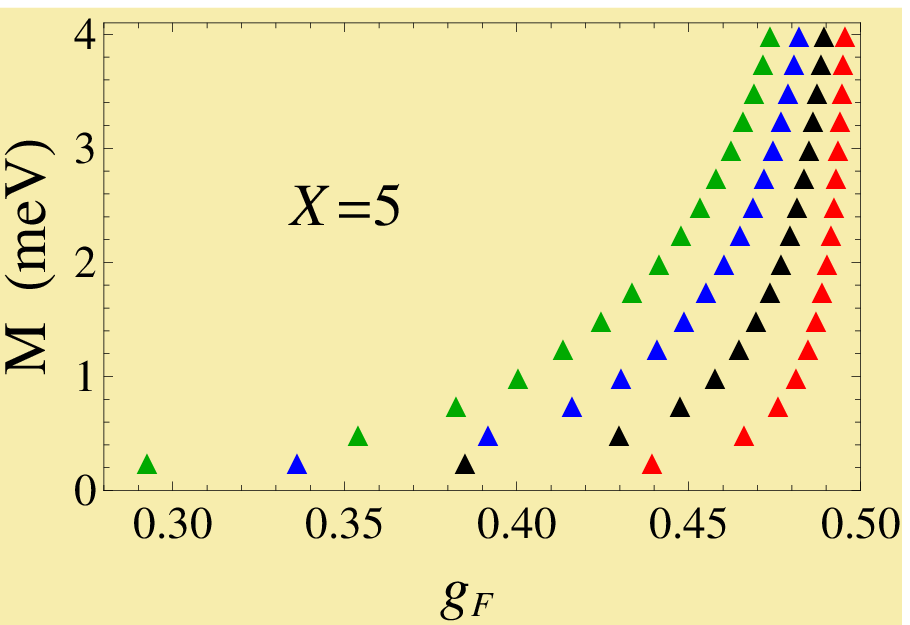}
\includegraphics[width=4.3cm,height=3.25cm]{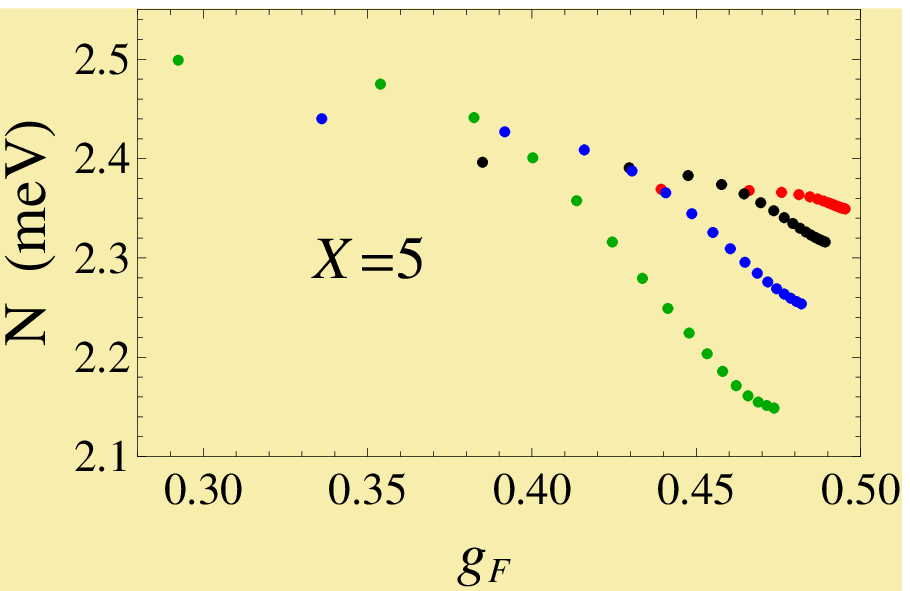}
\includegraphics[width=4.3cm,height=3.25cm]{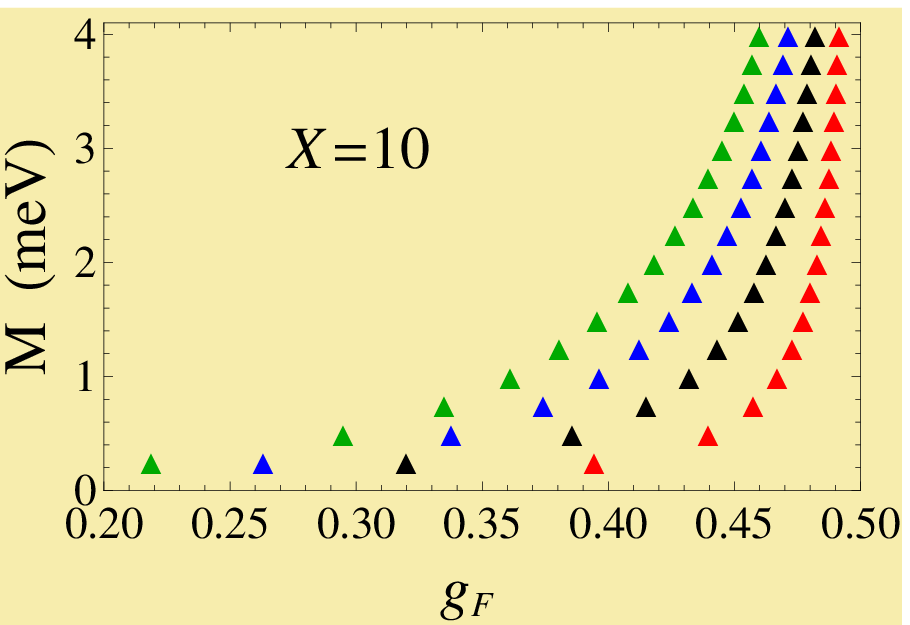}
\includegraphics[width=4.3cm,height=3.25cm]{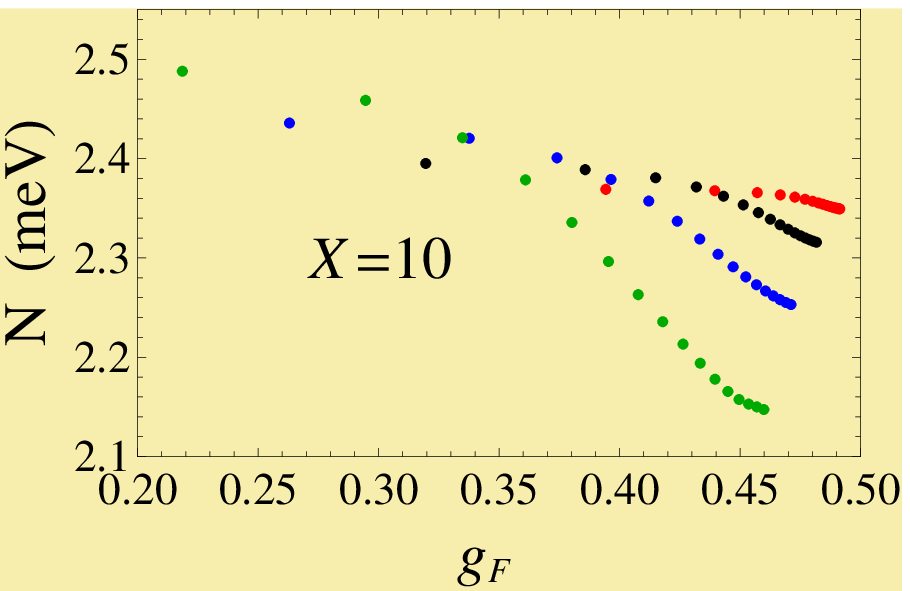}
\caption[] {(Color online) Self consistent solution of the FM (M) and LAF(N) order parameters, in tilted magnetic fields, when $B_\parallel =X B_\perp$, with $B_\perp=0.1$T(red), $0.2$T(black), $0.3$T(blue), and $0.4$T(green), as functions of ferromagnetic interaction $g_F$.}\label{fitexperiment}
\end{figure*}

I here provide some additional numerical results for the self-consistent solution of the LAF and FM orders, when the BLG is subject to a tilted magnetic field. All together, the results presented in the main part of the paper, and the ones shown in Fig.~1, may strengthen my claim that there exists no direct transition between the easy-plane LAF state to a pure FM state, as one increases the the parallel component of the magnetic field, while keeping its perpendicular component fixed.


\begin{thebibliography}{99}
\bibitem{novoselov} K. S. Novoselov, A. K. Geim, S. V. Morozov, D. Jiang, M. I. Katsnelson, I. V. Grigorieva, S. V. Dubonos, and A. A. Firsov, Nature (London) {\bf 438}, 197 (2005). 
\bibitem{kimfirstpaper} Y. Zhang, Y.-W. Tan, H. L. Stormer, and P. Kim, Nature (London) {\bf 438}, 201 (2005).
\bibitem{sharapov} V. P. Gusynin, and S. G. Sharapov, Phys. Rev. Lett. {\bf 95}, 146801 (2005).
\bibitem{falko} E. McCann, and V. I. Fal'ko, Phys. Rev. Lett. {\bf 96}, 086805 (2006).
\bibitem{yacoby} R. T. Weitz,  M. T. Allen, B. E. Feldman, J. Martin, A. Jacoby, Science {\bf 330}, 812 (2010).
\bibitem{weiss-PRL} F. Freitag, J. Trbovic, M. Weiss, C. Sch$\ddot{\mbox{o}}$nenberger, Phys. Rev. Lett. {\bf 108}, 076602 (2012).
\bibitem{Jairo} J. Velasco Jr., L. Jing, W. Bao, Y. Lee, P. Kratz, V. Aji, M. Bockrath, C.N. Lau, C. Varma, R. Stillwell, D. Smirnov, Fan Zhang, J. Jung, A.H. MacDonald, Nat. Nano. {\bf 7}, 156 (2012).
\bibitem{titltedfieldBLG} F. Freitag, W. Weiss, R. Maurand, J. Trbovic, C. Sch$\ddot{\mbox{o}}$nenberger, Phys. Rev. B {\bf 87}, 161402(R) (2013). 
\bibitem{weiss-solstatecommun} F. Freitag, M. Weiss, R. Maurand, J. Trbovic, and C. Sch$\ddot{\mbox{o}}$nenberger, Solid State Commun. {\bf 152}, 2053 (2012).
\bibitem{bitanclassification} B. Roy, Phys. Rev. B {\bf 88}, 075415 (2013).
\bibitem{nandkishore-levitov} R. Nandkishore, and L. Levitov, Phys. Rev. Lett. {\bf 104}, 156803 (2010) ; arxiv:1002.1966. 
\bibitem{kun-oskar} O. Vafek and K. Yang, Phys. Rev. B {\bf 81}, 041401(R) (2010).
\bibitem{oskar} O. Vafek, Phys. Rev. B {\bf 82}, 205106(2010).
\bibitem{robert} R. E. Throckmorton and O. Vafek, Phys. Rev. B {\bf 86}, 115447 (2012).
\bibitem{Kveshchenko-MLG} D. V. Kveshchenko, Phys. Rev. Lett. {\bf 87}, 246802 (2001).
\bibitem{miransky-MLG-QHE} V.P. Gusynin, V.A. Miransky, S.G. Sharapov, I.A. Shovkovy, Phys. Rev. B {\bf 76}, 195429 (2006). 
\bibitem{herbut-originalQHE} I. F. Herbut, Phys. Rev. B {\bf 75}, 165411 (2007).
\bibitem{bitan-oddQHE} B. Roy, Phys. Rev. B {\bf 84}, 035458 (2011).
\bibitem{miransky} E. V. Gorbar, V. P. Gusynin, and V. A. Miransky, Phys. Rev. B {\bf 81}, 155451 (2010); E. V. Gorbar, V. P. Gusynin, J. Jia and V. A. Miransky, \emph{ibid} {\bf 84}, 235449 (2012); E. V. Gorbar, V. P. Gusynin, V. A. Miransky, and I. A. Shovkovy, \emph{ibid}. {\bf 85}, 235460 (2012).
\bibitem{prange} For quantum Hall physics in regular 2DEG see \emph{Quantum Hall effect} 2nd edition, edited by R. E. Prange and S. M. Girvin, (Springer-Verlag, New York, 1989). 
\bibitem{herbutso3} I. F. Herbut, Phys. Rev. B {\bf 76}, 085432 (2007).
\bibitem{Mcdonald} F. Zhang, H. Min, M. Polini, A.H. MacDonald, Phys. Rev. B, {\bf 81}, 041402 (R) (2010).
\bibitem{yacoby-scaling} J. Martin, B. E.  Feldman, R. T. Weitz, M. T. Allen, and A. Yacoby, Phys. Rev. Lett. {\bf 105}, 256806 (2010).
\bibitem{kim-2010-nu2} Y. Zhao, P. Cadden-Zimansky, Z. Jiang, and P. Kim, Phys. Rev. Lett. {\bf 104}, 066801 (2010).
\bibitem{velasconu2} J. Velasco Jr., Y. Lee, Z. Zhao, L. Jing, P. Kratz, M. Bockrath, C. N. Lau, arxiv:1303.3649.
\bibitem{herbut-book} I. Herbut \emph{A Modern Approach to Critical Phenomena}, (Cambridge University Press, Cambridge, 2007).
\bibitem{herbut-juricic-roy} I. F. Herbut, V. Juri\v ci\'c, and B. Roy, Phys. Rev. B {\bf 79}, 085116 (2009); B. Roy, and I. F. Herbut, Phys. Rev. B {\bf 82}, 035429 (2010).
\bibitem{supplementary} See Supplementary Material for detail of LL wave-functions, spectrum of $H_{HF}$; the variational calculation of gap equations, exciation spectrum; and additional numerical results.
\bibitem{finitesizeBLG} C. Yannouleas, I. Romanovsky, and U. Landman, Phys. Rev. B {\bf 82}, 125419 (2010).
\bibitem{cutoffcoment} Cut-off in a four-band model for BLG is $\sim 3$eV.  
\bibitem{kharitonov} M. Kharitonov, Phys. Rev. B {\bf 86}, 195435 (2012).
\bibitem{Hubbardmodel} Within the framework of continuum description of the Hubbard model in BLG $V_0=V_{2K}$, and thus $\tilde{V}=0$ at the scale $\Lambda \sim 0.2$eV.  
\bibitem{miransky-MLG-QHE-2} V.P. Gusynin, V.A. Miransky, S.G. Sharapov, I.A. Shovkovy, Phys. Rev. B {\bf 77}, 205409 (2008). 
\bibitem{abanin} D. A. Abanin, K. S. Novoselov, U. Zeitler, P. A. Lee, A. K. Geim, L. S. Levitov, Phys. Rev. Lett. {\bf 98}, 196806 (2007).
\bibitem{kim-patric} P. Maher, C. R. Dean, A. F. Young, T. Taniguchi, K. Watanabe, K. L. Shepard, J. Hone, and P. Kim, Nat. Phys. {\bf 9}, 154 (2013). 
\bibitem{bitanscaling} I. F. Herbut, and B. Roy, Phys. Rev. B {\bf 77}, 245438 (2008),  B. Roy, and I. F. Herbut, \emph{ibid} {\bf 83}, 195422 (2011); \emph{ibid} {\bf 88}, 045425 (2013).
\bibitem{notationhopping} In literature sometime different notations are used for various hopping parameters, where $t_0 \to \gamma_0$, $t_\perp \to \gamma_1$, and $t_1 \to \gamma_4$. See for example Ref.\onlinecite{bandparameter}.
\bibitem{bandparameter} L. M. Zhang, Z. Q. Li, D. N. Basov, M. M. Fogler, Z. Hao, and M. C. Martin, Phys. Rev. B {\bf 78}, 235408 (2008).
\bibitem{barlas} R. Cote, J. Lambert, Y. Barlas, and A. H. MacDonald, Phys. Rev. B {\bf 82}, 035445 (2010).
\bibitem{khveshchenko} D. V. Khveshchenko, Phys Rev. B {\bf 75}, 153405 (2007).
\bibitem{jain} J. K. Jain, \emph{ Composite Fermions}, (New York: Cambridge University Press, 2007). 
\bibitem{laufractional} W. Bao, Z. Zhao, H. Zhang, G. Liu, P. Kratz, L. Jing, J. Velasco Jr., D. Smirnov, C. N. Lau, Phys. Rev. Lett. {\bf 105}, 246601 (2010).
\end{thebibliography}

\begin{thebibliography}{99}
\bibitem{robertsupple} R. E. Throckmorton, and O. Vafek, Phys. Rev. B {\bf 86}, 115447 (2012).
\bibitem{LauJairo} J. Velasco Jr., L. Jing, W. Bao, Y. Lee, P. Kratz, V. Aji, M. Bockrath, C.N. Lau, C. Varma, R. Stillwell, D. Smirnov, Fan Zhang, J. Jung, A.H. MacDonald, Nat. Nano. {\bf 7}, 156 (2012). 
\bibitem{baselSSC} F. Freitag, M. Weiss, R. Maurand, J. Trbovic, and C. Sch$\ddot{\mbox{o}}$nenberger, Solid State Commun. {\bf 152}, 2053 (2012).
\bibitem{baselRC} F. Freitag, W. Weiss, R. Maurand, J. Trbovic, and C. Sch$\ddot{\mbox{o}}$nenberger, Phys. Rev. B {\bf 87}, 161402(R) (2013).
\bibitem{vafekRGpaper} O. Vafek, Phys. Rev. B {\bf 82}, 205106 (2010).
\bibitem{herbutso3AF} I. F. Herbut, Phys. Rev. B {\bf 76}, 085432 (2007).
\end{thebibliography}
\end{document}